\documentclass{article}

\usepackage[utf8]{inputenc}
\usepackage[T1]{fontenc}
\usepackage{hyperref}
\usepackage{amsmath, amsfonts, amsthm}
\usepackage{graphicx}
\usepackage{url}

\usepackage{hyperref}  

\title{\huge{Rethinking AI in the age of climate collapse: Ethics, power, and responsibility}}
\author{Julio Vega\thanks{julio.vega@urjc.es\\Department of Telematic Systems and Computing, Rey Juan Carlos University, Camino del Molino, n.º 5, 28942 - Fuenlabrada (Madrid)}}
\date{\today}

\begin{document}

\maketitle

\section*{Abstract}

The climate crisis requires responses that integrate scientific, ethical, social, and technological perspectives. Artificial intelligence (AI) has emerged as a powerful tool in climate modelling, environmental monitoring, and energy optimisation, yet its growing use also raises critical environmental, ethical, legal, and social questions. This contribution examines the ambivalent role of AI in the ecological crisis, addressing both its promises and its risks. On the one hand, AI supports improvements in climate forecasting, renewable energy management, and real-time detection of environmental degradation. On the other hand, the energy demands of data centres, resource-intensive hardware production, algorithmic bias, corporate concentration of power, and technocratic decision-making reveal contradictions that challenge its sustainability. The discussion explores these issues through interdisciplinary lenses, including environmental ethics, philosophy of technology, and legal governance, and concludes with recommendations for socially just, ecologically responsible, and democratically accountable uses of AI. Rather than assuming AI as an inherently sustainable solution, this analysis argues that its contribution to climate action depends fundamentally on the values, institutions, and power structures that shape its development.

\emph{Keywords: Artificial intelligence; Climate crisis; Environmental ethics; Sustainability; Algorithmic governance; Environmental justice; Philosophy of technology; Green AI; Regulation}

\section{Introduction}

The climate crisis stands as one of the greatest challenges of the twenty-first century. Its complexity demands responses that integrate scientific, ethical, social, and technological forms of knowledge. In this context, artificial intelligence (AI) emerges as a tool capable of transforming the ways in which we analyse, predict, and address the impacts of climate change. Yet far from providing a simple solution, this technology activates ethical, environmental, and social debates that require critical and interdisciplinary reflection.

This study offers an analytical and deliberately reflective examination of the role of AI in the contemporary ecological crisis. The discussion spans philosophical, ethical, legal, and social questions, aiming to problematise the growing dependence on algorithmic systems in environmental governance. Although AI has been promoted as a driver of efficiency and sustainability, it also raises risks related to intensifying energy consumption, reproducing inequality, and consolidating technocratic approaches to environmental management.

From a philosophical perspective, we must ask how AI reshapes our relationship with nature: does it function as a means toward more sustainable models, or does it perpetuate an instrumental view of the natural world? Ethically, the delegation of sensitive decisions to opaque systems requires examining transparency and fairness. Legally, the absence of robust regulatory frameworks opens the door to uncontrolled uses; and socially, AI’s uneven development generates tensions concerning who benefits and who bears the costs of its deployment.

The paper is structured into eight sections. It begins with an overview of AI’s role in climate mitigation and adaptation, followed by an analysis of its direct environmental impacts. The discussion then addresses the ethical and philosophical dilemmas associated with AI deployment, examines the regulatory frameworks required to govern these technologies, and considers the social implications arising from their use. The concluding section synthesises the findings and provides recommendations for fostering AI practices that support sustainability, environmental justice, and equitable governance.

\section{AI as a Tool in the Fight Against Climate Change}

Artificial intelligence has become a key component of the digital transformation of various sectors, including the environmental sphere. Its capacity to process large volumes of data and generate predictions from complex patterns has expanded our understanding of climate change and strengthened mitigation and adaptation strategies. Its contributions include improvements to climate modelling, energy optimisation, and real-time environmental monitoring.

In the scientific domain, these technologies have enabled advances in simulating atmospheric and oceanic behaviour using machine learning techniques and deep neural networks. These tools, used by bodies such as the IPCC, facilitate the production of more precise and dynamic future scenarios. At the same time, automated analysis of satellite imagery enables the detection of deforestation, fires, or changes in land cover, allowing rapid responses and improving environmental enforcement.

AI has also become relevant in energy management, optimising the production, distribution, and consumption of renewable energy \cite{venkataswamy2022rare, sarkar2024carbon}. Intelligent systems can anticipate demand, adjust electrical grids, and support the integration of solar and wind energy. Such advances are often celebrated as part of a more efficient transition away from fossil fuels.

However, this optimistic picture must be tempered by a critical perspective. The development and application of such technologies is concentrated largely in the Global North, reproducing structural inequalities and generating significant disparities in technical capacity. Simultaneously, AI may be used as a tool for \emph{greenwashing}, offering superficial solutions that conceal the lack of deeper changes in production models.

The transformative potential of AI will ultimately depend on the principles guiding its implementation. For AI to meaningfully contribute to sustainability, it must be oriented by ethical criteria and principles of environmental justice, avoiding unequal distribution of benefits and preventing reinforcement of extractive ecological dynamics \cite{grigoryan2018grasp}.

\section{Environmental Impact of AI}

Despite its potential contributions, AI generates significant environmental impacts that require rigorous scrutiny. Many algorithmic solutions rely on infrastructures that demand large amounts of energy and natural resources, potentially exacerbating the very ecological crisis they aim to mitigate. This paradox invites a critical assessment of the sustainability of the technological ecosystem itself.

One major issue is the high energy consumption of data centres that host training and processing infrastructures. The development of advanced deep learning models requires intensive computational capacity, translating into substantial electricity use. Some estimates suggest that training a single large language model can produce more than 280 tonnes of CO\textsubscript{2}, a figure comparable to the lifetime emissions of multiple cars.

In addition, water use for cooling servers poses a serious concern, especially in regions already suffering hydrological stress. Locating technological infrastructure in vulnerable areas can exacerbate socio-environmental tensions, affecting ecosystems and local communities.

Environmental impacts also extend to the hardware involved. Producing chips, servers, and intelligent devices requires extracting rare earth minerals and heavy metals, activities often associated with pollution, biodiversity loss, and labour rights violations. The lifecycle of this equipment also poses challenges concerning electronic waste management and planned obsolescence.

These tensions reveal a deep ethical contradiction: can a technology be considered sustainable if its development contributes to environmental degradation? Philosophically, this paradox echoes critiques of the instrumental rationality guiding much of modern technological development.

In response, initiatives such as \emph{Green AI} promote energy efficiency, renewable energy use, and algorithms that demand fewer resources \cite{schwartz2019green}. Although promising, these proposals remain marginal relative to the rapid growth of the technological sector. Their effectiveness will depend on regulatory frameworks that incentivise responsible practices and limit excessive resource use.

\section{Ethical Issues in the Application of AI to Environmental Governance}

The use of AI in environmental management raises far-reaching ethical dilemmas. As decisions are increasingly delegated to algorithmic systems, questions arise regarding responsibility, transparency, justice, and democratic control. These are not merely technical issues but central elements in evaluating the legitimacy of such technologies.

The opacity of many deep learning models is a primary challenge. These “black boxes” make it difficult to understand the rationale behind decisions that may affect entire ecosystems or communities. Lack of explainability can undermine basic principles of procedural justice.

Further challenges stem from the biases embedded in training data. Far from neutral, AI systems can reproduce existing inequalities and favour solutions aligned with urban, corporate, or Global North interests, while neglecting the needs of rural, Indigenous, or marginalised groups. This risk is particularly concerning in the climate context, where environmental justice is essential.

Concentration of power in large technology companies raises additional ethical concerns. The privatisation of knowledge and AI infrastructures can sideline democratic deliberation and limit local communities’ capacity to influence environmental decisions that directly affect them. This raises questions about democratic limits and the influence of corporate actors on climate policy.

The lure of \emph{technological solutionism} is another ethical pitfall. Presenting the climate crisis as an optimisable problem can obscure its structural and ethical roots. A critical perspective reminds us that climate change is fundamentally a socio-economic and political issue, not reducible to algorithmic efficiency.

Environmental ethics therefore requires transparency, inclusiveness, and accountability in the design and deployment of AI. Systems must be auditable, data must be representative, and affected communities should be meaningfully involved in decision-making. AI for sustainability cannot replace social justice; it must accompany and reinforce it.

\section{Philosophical Perspectives on AI and Nature}

The introduction of AI into environmental governance reopens philosophical debates about the relationship between humanity, technology, and nature. Is AI a path toward new forms of coexistence with the planet, or an extension of the extractivist logic that has characterised modernity?

From a Cartesian standpoint, nature has long been viewed as an object available for human manipulation. AI could be seen as the culmination of this instrumental perspective. However, alternative frameworks from care ethics, deep ecology, and posthumanism offer other interpretations. Care ethics emphasises interdependence and shared vulnerability; deep ecology foregrounds the intrinsic value of all living beings; and posthumanism invites us to consider AI as a new agent within ecological assemblages, reshaping notions of responsibility.

The debate over technological neutrality is also relevant. Following authors such as Winner \cite{winner1986whale}, technologies embed values and power structures. AI reflects the priorities of its designers \cite{mueller2023mathematical}: efficiency, economic profit, social welfare, or ecological sustainability.

Finally, tensions arise between the temporalities of AI—driven by immediate optimisation—and those of ecology, which require precaution and long-term thinking. Designing technologies that respect non-human rhythms and that incorporate the precautionary principle is one of the deepest challenges at stake.

\section{Legal and Regulatory Frameworks for AI in Environmental Contexts}

Regulation of AI in environmental applications remains in an early stage, marked by legal gaps and by insufficient integration between technological and climate policy \cite{friendsoftheearth2025ai}. Although the EU has advanced proposals such as the AI Act \cite{eu2021ai}, these frameworks focus on general risks and do not adequately address ecological or distributive impacts.

Internationally, principles such as precaution and sustainable development offer valuable normative foundations, yet lack binding mechanisms concerning AI. Countries in the Global South face additional disadvantages: limited regulatory capacity and pressures to adopt technologies developed elsewhere \cite{blair2024ai}.

Legal responsibility presents a major unresolved issue. Determining who is accountable for environmental harm caused by algorithmic systems—developers, operators, or public authorities—remains unclear \cite{ejatlas2024conflictos}. Establishing clear legal frameworks, along with independent oversight bodies, is essential to ensure just governance of these technologies.

\section{Social Implications of AI in the Climate Crisis}

The social implications of AI in climate governance are deep and often ambivalent. The technological divide between the Global North and Global South, potential automation of labour sectors, and the risk of culturally or politically biased environmental surveillance may exacerbate existing inequalities \cite{wmo2025aiwarnings}.

The concentration of technical capacities in a few major actors consolidates new forms of power, while many vulnerable communities—those most affected by the climate crisis—are excluded from the design and evaluation of these systems. Without policies for a just transition, automation linked to green economies may generate legitimate resistance.

Nevertheless, positive experiences show that AI can also be appropriated by communities and social movements \cite{tiltack2025aejim, proma2021cleanairnowkc}. Early warning systems, open-data platforms, and pollution-mapping tools have strengthened environmental justice initiatives and expanded citizen participation.

\section{Conclusions and Recommendations}

The analysis of AI’s role in the climate crisis reveals a complex and deeply ambivalent landscape. AI offers significant opportunities to improve climate modelling, optimise resources, and monitor environmental risks, yet it also produces ecological, ethical, and social impacts that must not be ignored.

From an interdisciplinary perspective, several recommendations emerge:

\begin{enumerate}
\item Establish public policies regulating the use of AI according to ecological, ethical, and social criteria, incorporating the principles of precaution, environmental justice, and human rights.
\item Ensure transparency and auditability of systems used in environmental decision-making, strengthening mechanisms of accountability.
\item Promote inclusive AI that integrates community knowledge and local expertise, avoiding decontextualised technocratic solutions.
\item Develop international standards concerning the environmental impact of algorithmic infrastructures, incentivising the use of renewable energy.
\item Advance ethical and ecological training in engineering and data science programmes, fostering a professional culture oriented toward sustainability.
\end{enumerate}

In sum, artificial intelligence can be an ally in ecological transition, provided it is governed by principles of equity, sustainability, and democratic responsibility. The question is no longer whether we should use AI to confront the climate crisis, but how to orient it so that it serves the common good and the flourishing of life in all its forms.

\section*{Funding}

Not Applicable.

\section*{Competing Interest} 

Not Applicable.

\section*{Data Availability}

Not Applicable.

\clearpage


\begin{thebibliography}{99}

\bibitem{schwartz2019green}
R. Schwartz, J. Dodge, N. A. Smith, O. Etzioni,
\textit{Green AI},
arXiv preprint arXiv:1907.10597, 2019,
Available at: \url{https://arxiv.org/abs/1907.10597}

\bibitem{venkataswamy2022rare}
V. Venkataswamy, J. Grigsby, A. Grimshaw, Y. Qi,
\textit{RARE: Renewable Energy Aware Resource Management in Datacenters},
arXiv preprint arXiv:2211.05346, 2022,
Available at: \url{https://arxiv.org/abs/2211.05346}

\bibitem{sarkar2024carbon}
S. Sarkar, S. Sharma, P. Kapadia, S. Ghosh, P. Balamuralidhar,
\textit{Carbon Footprint Reduction for Sustainable Data Centers in Real-Time},
arXiv preprint arXiv:2403.14092, 2024,
Available at: \url{https://arxiv.org/abs/2403.14092}

\bibitem{grigoryan2018grasp}
G. Grigoryan, J. Hines, V. Theodorou, D. Koutsonikolas, T. La Porta, G. Kesidis,
\textit{GRASP: A Green Energy Aware SDN Platform},
arXiv preprint arXiv:1804.09542, 2018,
Available at: \url{https://arxiv.org/abs/1804.09542}

\bibitem{winner1986whale}
L. Winner,
\textit{The Whale and the Reactor: A Search for Limits in an Age of High Technology},
University of Chicago Press, 1986.

\bibitem{mueller2023mathematical}
D. Müller, M. Chiodo,
\textit{Mathematical Artifacts Have Politics: The Journey from Examples to Embedded Ethics},
arXiv preprint arXiv:2308.04871, 2023,
Available at: \url{https://arxiv.org/abs/2308.04871}

\bibitem{eu2021ai}
European Union,
\textit{Proposal for a Regulation of the European Parliament and of the Council Laying Down Harmonised Rules on Artificial Intelligence (Artificial Intelligence Act) and Amending Certain Union Legislative Acts},
COM(2021) 206 final, 2021,
Available at: \url{https://eur-lex.europa.eu/legal-content/EN/TXT/?uri=COM:2021:206:FIN}

\bibitem{blair2024ai}
Tony Blair Institute for Global Change,
\textit{How Leaders in the Global South Can Devise AI Regulation That Enables Innovation}, 2024,
Available at: \url{https://institute.global/insights/tech-and-digitalisation/how-leaders-in-global-south-can-devise-ai-regulation-that-enables-innovation}

\bibitem{survival2024unesco}
Survival International,
\textit{UNESCO Promotes a Conservation Model That Destroys Indigenous Peoples}, 2024,
Available at: \url{https://www.survival.es/informes/informe-decolonizeunesco}

\bibitem{tiltack2025aejim}
T. Tiltack,
\textit{AEJIM: A Real-Time AI Framework for Crowdsourced, Transparent, and Ethical Environmental Hazard Detection and Reporting},
arXiv preprint arXiv:2503.17401, 2025,
Available at: \url{https://arxiv.org/abs/2503.17401}

\bibitem{proma2021cleanairnowkc}
R. A. Proma et al.,
\textit{CleanAirNowKC: Building Community Power by Improving Data Accessibility},
arXiv preprint arXiv:2107.11633, 2021,
Available at: \url{https://arxiv.org/abs/2107.11633}

\bibitem{friendsoftheearth2025ai}
Friends of the Earth,
\textit{Harnessing AI for Environmental Justice}, 2025,
Available at: \url{https://policy.friendsoftheearth.uk/reports/harnessing-ai-environmental-justice}

\bibitem{wmo2025aiwarnings}
World Meteorological Organization,
\textit{AI for Early Warnings: Experts Unite with Calls to Build Resilience and Protect Development Gains}, 2025,
Available at: \url{https://wmo.int/media/news/ai-early-warnings-experts-unite-calls-build-resilience-and-protect-development-gains}

\bibitem{ejatlas2024conflictos}
EJAtlas,
\textit{Environmental Justice Atlas}, 2024,
Available at: \url{https://ejatlas.org/}

\end{thebibliography}
\end{document}